\documentclass{appolb}
\input epsf.tex

\def\ni{\noindent}
\def\ms{\medskip}

\newcount\rys

\def\rysunek#1#2{
\global\advance\rys by 1 \vbox{\centerline{\epsfxsize=4.5truein\epsfbox{#1}}
\vskip 0.1truein
\centerline{\vtop{\ni Fig. \the\rys . #2}}}
\vskip 0.2truein minus 0.2truein}

\begin{document}

\title{ STRUCTURE OF PROTON COMPONENT OF NEUTRON \\
STAR MATTER FOR REALISTIC NUCLEAR MODELS }

\author{ M. Kutschera$^{a,b}$, S. Stachniewicz$^a$,\\
 A. Szmagli\'nski$^c$ and W. W\'ojcik$^c$
\address{$^a$~H. Niewodnicza\'nski Institute of Nuclear Physics,\\
ul. Radzikowskiego 152, 31-342 Krak\'ow, Poland \\ 
$^b$~Institute of Physics, Jagiellonian University,\\
ul. Reymonta 4, 30-059 Krak\'ow, Poland\\
$^c$~Institute of Physics, Technical University,\\
ul. Podchor\c{a}\.zych 1, 30-084 Krak\'ow, Poland }
}

\maketitle

\begin{abstract} 

We study  properties of the proton component of neutron star matter   
for a number of realistic nuclear models. 
Protons which form a few percent admixture tend to be localized in potential 
wells corresponding to neutron matter 
inhomogeneities created by the protons in the neutron medium. We calculate the 
energy of the Wigner-Seitz cell enclosing a single localized proton. 
The neutron background is treated in the Thomas-Fermi approximation and 
the localized proton is described by the Gaussian wave function. The 
neutron density 
profile is obtained by solving the appropriate variational equation. 
This approach gives lower energies of localized protons than obtained 
previously 
with less sophisticated methods.

\end{abstract}


\section{Introduction}

Physical properties of the neutron star interior relevant to 
macroscopic observables are rather sensitive to the
the microscopic structure of dense nuclear matter in neutron stars. 
For example, the transport and magnetic 
properties of  neutron stars  depend strongly on the structure  of the so 
called liquid
core. Particularly important is the structure of the proton component. A 
uniform proton distribution and a  periodic
(crystalline) proton arrangement result in very different properties 
\cite{KuPLB}. The latter possibility was discussed in 
Refs. \cite{KuWoAPPB21,KuWoPL} for
strongly asymmetric nuclear matter which was shown to be unstable with respect
to proton localization. The localization effect is a result of the
interaction of protons with small density oscillations of the neutron
background \cite{KuWoPR}. The protons behave as localized polarons which form 
a periodic lattice at high densities \cite{KuWoNP}.

The presence of the localized protons inside neutron star cores would have
profound astrophysical consequences. The transport properties of such a phase
are rather different from that of a uniform nuclear matter \cite{BaHaAA}. In
particular, the cooling proceeds in a quite different way. Recent analysis 
\cite {BaHaAA} shows that the presence of such localized proton phase results
in more satisfactory fits of temperatures of observed neutron stars. 
Also, spin ordering of localized protons could strongly affect magnetic 
properties 
of the system \cite{KuWoAPPB21,KuWoAPPB27}. The spin ordered phase can 
contribute significantly to the
observed magnetic moments of neutron stars \cite{KuWoAPPB23,KuMNRAS}.

The aim of this paper is to study the proton localization for a number of 
realistic
nuclear models with improved variational method. In original calculations 
\cite{KuWoPL,KuWoPR}
both the proton wave function and the neutron density distribution were
assumed to be trial functions which included variational parameters.  In
this paper we find better estimates of energies of localized protons by solving
the appropriate variational equation for the neutron density profile that gives
the minimum energy for a fixed wave function of the localized proton.

The paper is organized as follows: In the next section we describe 
the model of proton impurities in the neutron star matter. 
In Sect.3 simple estimates of the proton localization based on trial 
functions 
are given. Sect.4 contains the formulation and the solution of the variational 
problem. Results are discussed in Sect.5.

\section{Model of proton impurities in the neutron star matter}

The amount of protons present in the neutron star matter, which is charge 
neutral 
and $\beta$-stable, is crucial for the cooling rate of neutron 
stars and also plays an important role for magnetic 
and transport properties of neutron star matter. 
Nuclear models do not uniquely predict the proton fraction 
of the neutron star matter at high densities. This controversy is discussed 
in details in Ref. \cite{KuPLB,KuZP} where the discrepancy of the proton 
fraction 
in various models is shown to reflect the uncertainty  
of the nuclear symmetry energy at high densities. In this paper we consider 
a class of nuclear interaction models for which the proton fraction is of the 
order of a few percent 
and decreases at high densities - as shown in Fig.1.
For the calculations we have chosen six realistic nuclear interaction models. 
These are interactions derived by  
Myers and Swiatecki \cite{MySwZSP}(MS), the Skyrme potential with parameters 
from Ref. 
\cite{VaBrPL,RaBePePRL} (Sk), 
the Friedman and Pandharipande interactions \cite{FrPaNP} (as parametrized by 
Ravenhall in 
Ref.\cite{LaARNPS}) (FPR)
and three models, UV14+TNI, AV14+UVII and UV14+UVII, from 
Ref.\cite{WiFiFaPR} by Wiringa et al.

Let us consider a neutron star matter containing a small proton fraction $x$. 
To compare the energy of a normal phase of uniform density 
and a phase with localized protons we apply
the Wigner-Seitz approximation and divide the system into cells, 
each of them enclosing a single proton  \cite{KuWoAPPB21,KuWoPR}. 
For simplicity, the cells are assumed to be spherical. The volume of the cell 
is $V={1/n_P}$. The normal phase is of uniform density $n_N$ 
and the neutron chemical potential is $\mu _N$. In the uniform density 
phase
protons are not localized and their wave functions are plane waves.

The energy of the cell, which is a sum of proton and neutron energies, reads

\begin{equation}
E_0 =V\epsilon \left( n_N,n_P \right),  \label{Cell Energy}
\end{equation}

\ni where $\epsilon \left( n_N,n_P\right) $ is the energy density of the 
uniform phase. 
For small proton density, i.e. for low $x$, we can expand the energy density

\begin{equation}
\epsilon \left( n_N,n_P\right) \approx \epsilon \left( n_N,0\right) 
+\mu_P\left( n_N,0\right) n_P. \label{Energy Density}
\end{equation}

In the following we shall adopt abbreviations 
$\epsilon \left( n_N\right)=\epsilon \left( n_N,0\right)$ for the energy 
density 
of pure neutron matter and $\mu_P\left( n_N\right)=\mu_P\left( n_N,0\right)$ 
for the proton chemical potential in pure neutron matter. The energy of the 
cell 
is thus approximately

\begin{equation}
E_0=\mu _P\left( n_N\right) +V\epsilon \left( n_N\right).  \label{Cell Energy2}
\end{equation}

Our aim is to compare the energy of the normal phase, where protons are 
nonlocalized, 
with the energy of a phase where the protons are trapped into potential wells, 
corresponding to the nonuniform neutron density distribution, which most 
likely 
form a regular arrangement. We treat this proton "crystal" 
in the Wigner-Seitz approximation.

Let us consider a Wigner-Seitz cell with nonuniform neutron matter 
distribution 
$n\left( r\right)$ surrounding the proton whose wave function is $\Psi _P$. 
In the local density approximation one can identify the proton effective 
potential 
with the local proton chemical potential $\mu _P\left( n\right)$ 
\cite{KuWoAPPB21}. 
The proton's effective potential varies locally with neutron matter density 
$n\left( r\right)$. This results in a potential well 
$\mu _P\left( n\left( r\right) \right)$ which affects the single proton wave 
function. 
The energy of the Wigner-Seitz cell, $E_L$, is:

$$ 
E_L=\int_{V}\{\Psi_P^{*}\left( r\right) \left[ -{\nabla ^2 \over 2m_P} 
+ \mu_P\left( n\left( r\right) \right) \right] \Psi_P\left( r\right) $$
\begin{equation}
+\epsilon \left( n\left( r\right) \right) + B_N\left( \overrightarrow{\nabla }
n\left( r\right) \right) ^2 \} \ d^3r . 
\label{EnLocKin}
\end{equation}

The first term is the energy of the proton in the effective potential 
$v_{eff} \left( r\right) =\mu _P\left( n\left( r\right) \right)$. 
It is by construction the attractive potential well. 
At high densities the derivative of the proton chemical potential is positive, 
${\partial \mu_P \over \partial n}>0$,
for all interactions we use. This can be seen in Fig.2 where the proton 
chemical potential in pure neutron matter is shown for nuclear interaction 
models from Fig.1. The neutron density profile
$n\left( r\right)$ 
is thus assumed to have a minimum at the center of the cell.

The two other terms in eq.(\ref{EnLocKin}) describe the neutron background 
contributions to the energy. These represent the neutron Fermi sea energy 
and the curvature energy due to the gradient of the neutron distribution, 
respectively, in the Thomas-Fermi approximation. 
Here $\epsilon \left( n\left( r\right) \right)$ is the local neutron matter 
energy 
per unit volume. The parameter $B_N$ is the curvature coefficient for pure 
neutron matter \cite{KuWoAPPB21}.

To decide which is the ground state configuration we compare the energies 
$E_0$ and $E_L$ assuming the same number of neutrons in the cell. 
This means that the neutron density variation conserves the baryon number:

\begin{equation}
\int_{V} \left( n\left( r\right) - n_N\right) \ d^3r = 0 . 
\end{equation}

\ni In the next section the minimum of the energy difference $\Delta E= 
E_L-E_0$   
is calculated in a simple variational approach and in Sect.4 
more sophisticated method is developed.

\section{Simple estimate of the localized proton energy}

We assume a simple trial form of the proton wave function and 
the neutron density variation. For the proton wave function we use a Gaussian 
form:

\begin{equation}
\Psi _P\left( r \right) =\left( {2 \over 3}\pi R_P^2\right)^
{-{ 3 \over 4}} 
\exp \left( -{ 3 \over 4} {r^2 \over R_P^2} \right) .
\end{equation}

\ni Here $R_P$ is the rms radius of the localized proton probability 
distribution. 
We treat this quantity as a variational parameter and minimize 
the energy difference $\Delta E$ with respect to $R_P$.

Using the trial form of the proton wave function $\Psi _P\left( r \right)$ 
the energy difference $\Delta E$ becomes

$$ 
\Delta E={ 9 \over 8m_PR_P^2} 
+\int_{V} \{ \Psi_P^2 \left( r\right) \left( \mu_P \left( n\left(
r\right) \right) -\mu_P\left( n_N\right) \right) + $$ 
\begin{equation}
\epsilon \left( n\left( r\right) \right) -\epsilon \left( n_N\right) +
B_N\left( {dn\left( r\right) \over dr} \right)^2\} \ d^3r  .
\label{EnDiff}
\end{equation}

\ni The neutron density $n\left( r\right)$ is chosen to be 
\cite{KuWoAPPB21,KuWoPR}: 

\begin{equation}
n\left( r\right) =n_N+\alpha \left[ \Psi _P^{*}\left( r\right) \Psi _P\left(
r\right) - {1 \over V} \right] . \label{Neutron Dens}
\end{equation}

\ni Here $\alpha $ is the second variational parameter; $\alpha >0$ 
corresponds 
to the neutron density enhancement around the proton and $\alpha <0$ 
corresponds to the bubble in the neutron density near the proton.

We calculate the energy difference $\Delta E$, Eq.(\ref{EnDiff}), 
for small proton fraction $x$, i. e. in the limit of large volume $V$. 
The first and the last terms were calculated assuming that 
the Wigner-Seitz cell radius $R_C$ is much bigger than $R_P$, $R_C\gg R_P$. 
Denoting $\Psi _P^{*}\left( r\right) \Psi _P\left(r\right)=p\left( r\right)$ 
and expanding in $1 \over V$ we have

$$
\int_{V} p\left( r\right) \left( \mu _P\left( n_N+\alpha p\left(
r\right) -\alpha { 1 \over V} \right) -\mu _P\left( n_N\right) \right) 
d^3r = $$
$$
\int_{V} p\left( r\right) \left( \mu _P\left( n_N+\alpha
p\left( r\right) \right) -\mu _P\left( n_N\right) \right)  d^3r - $$  
\begin{equation}
 \alpha { 1 \over V} \int_{V} p\left( r\right) {\partial \mu _P \over 
\partial n} \left( n_N + \alpha p \left( r\right) \right)  d^3r .
\end{equation}

The integral in the last term does not depend on the cell volume 
so that this term vanishes in the limit $V\rightarrow \infty $. 
Expanding in the same way the energy density, we obtain from 
the third term in Eq.(\ref{EnDiff})

$$
\int_{V} \left[ \epsilon \left( n_N + \alpha p\left( r\right)
- \alpha { 1 \over V} \right) - \epsilon \left( n_N\right) \right]  
d^3r = $$
$$
\int_{V} \left[ \epsilon \left( n_N+\alpha p\left( r\right)
\right) -\epsilon \left( n_N\right) \right] \ d^3r - \alpha \mu _N\left( 
n_N\right) - $$
\begin{equation}
\alpha { 1 \over V} \int_{V} \left( \mu _N\left( n_N+\alpha p\left( r\right)
\right) -\mu _N\left( n_N\right) \right) \ d^3r .
\end{equation}

\ni Here also the integral in the last term does not depend on the cell 
volume, 
since $p\left( r\right) $ is a Gaussian, and this term vanishes for large $V$.
The last term containing the coefficient of curvature $B_N$ is easily 
evaluated to be:

\begin{equation}
\int_{V} B_N\left( \overrightarrow{\nabla }n\left( r\right) \right) ^2  d^3r =
{ 9 \over 2} \left( { 4 \over 3} \pi \right) ^{-{ 3 \over 2}}{ 1 \over R_P^5} 
B_N\alpha ^2 .
\end{equation}

\ni The energy difference $\Delta E$ thus becomes:

$$
\Delta E= { 9 \over 8m_PR_P^2}+ 
\int_{V} \left\{ \left( \mu _P\left( n\left( r\right)
\right) -\mu_P\left( n_N\right) \right) p\left( r\right) \right\} d^3r + $$
\begin{equation} 
\int_{V} \left\{ \epsilon \left( n\left( r\right)
\right) -\epsilon \left( n_N\right) \right\} \ d^3r - 
\alpha \mu _N\left( n_N\right)+ 
{ 9 \over 2} \left( { 4 \over 3} \pi \right)^{-{ 3 \over 2}} { 1 \over R_P^5} 
B_N\alpha ^2 . \label{EnDiff2}
\end{equation}

\ms

\ni We obtain physical parameters of the localized phase 
for a given neutron matter density $n_N$ by a straightforward minimization 
of $\Delta E$ with respect to the two variational parameters $\alpha $ and 
$R_P$.

The results of the calculations for the MS and FPR nuclear interactions 
are presented, respectively, in Figs.3 and 4 where we show the energy 
difference $\Delta E$ 
as a function of the proton distribution rms radius $R_P$ for a few values of 
the neutron density. The curves are 
labeled 
with the value of the neutron  matter density $n_N$ with subscript $\alpha $.
One can notice that for both MS and FPR interactions 
there appears a local minimum 
above a certain density, for the proton rms radius $R_P$ in the range 
$1fm-2fm$. We have chosen the results for MS and FPR interactions only as 
examples of a general behaviour which is observed for all interactions we use 
in the calculations (more detailed account of our calculations will be 
presented elsewhere).
With increasing neutron matter density $n_N$ the depth of the minimum 
increases 
and above the threshold density the energy difference becomes negative. 
The negative value $\Delta E<0$ means that the energy of the localized proton 
is lower than the energy of a nonlocalized proton and the localized proton 
state is preferred energetically.
The behaviour of $\Delta E$ is very similar 
for all interactions we examine.  This shows that the localization is not an 
effect 
of some specific interaction but rather is a general qualitative feature of 
the 
physical system we consider.
Quantitative results, i. e. the localization density, the value of $\Delta E$ 
at the minimum and the localization radius $R_P$, depend on the specific 
interactions. 
The proton localizaton occurs at the lowest density for the Skyrme 
interactions, $n_{loc}=0.4 fm^{-3}$,
and the energy difference $\Delta E$ displays the fastest decrease with the 
density.
One can say that the localization is the strongest 
in this case. 
 
To understand better the localization mechanism it seems usefull to consider 
separately various contributions to the total energy difference. 
In Figs.5 and 6 we show the proton contribution, $E_P$, to the energy 
difference $\Delta E$, 
which consists of kinetic and potential terms. Here the minimum occurs 
at lower values of the proton rms radius $R_P$. One should keep 
in mind that the proton energy contribution represents a difference 
of the kinetic and potential energies of a localized proton and a single 
plane-wave proton. In the latter case the kinetic energy is zero.
Next figures, Figs.7 and 8, show the contribution of the neutron background 
to the total energy difference. This contribution is a monotonically 
decreasing function of the proton rms radius $R_P$. It grows very fast for low 
values of 
$R_P$. 
This rapid growth is similar to the behaviour of the gradient term 
contribution 
which is displayed in Figs.9 and 10. Thus a sum of these contributions also 
grows fast for low values of $R_P$. Its values for a given radius $R_P$ 
increase 
with the mean neutron density $n_N$.

The minimum of the total energy difference, $\Delta E$, which is a sum 
of all contributions shown in Figs.5-10, is thus a result of a delicate 
balance between repulsive contributions due to the neutron background, the 
proton kinetic term and the neutron curvature energy, and 
the attractive part of the proton interaction energy. Results of our 
calculations for a number of effective nuclear interactions show that 
such a minimum occurs in all cases above some density. One may thus 
conclude that the localization of proton impurity in the neutron matter 
is a general prediction of nuclear models of the class we consider here.

The threshold density for proton localization, $n_{loc}$, depends also 
on the curvature coefficient $B_N$ entering the gradient term and 
the proton effective mass $m_P$, which are parameters of our model. 
In Fig.11 we show how $n_{loc}$ changes with $B_N$. One can notice that 
the localization density only weakly increases with the curvature coefficient 
in a wide range of its values. Also, the rms proton distribution radius 
at the threshold density $R_P^{loc}$ increases slowly with increasing 
curvature coefficient $B_N$, Fig.12. This fact is rather important in regard 
of validity of the Thomas-Fermi approach used in our model.

The threshold density $n_{loc}$ depends in a more sensitive way on the 
proton effective mass, $m_P$, as shown in Fig.13. For values of $m_P$ 
less than the bare proton mass the localization density increases. 
However, in the range $600MeV-938MeV$ which is most likely physically relevant 
to neutron stars, the threshold density changes by about $20\%$.

\section{The self-consistent method}

Variational calculations of the localized proton energy presented 
in the previous section used the trial functions with only two variational 
parameters, 
$\alpha $ and $R_P$. In this section we develop more advanced variational 
method 
which should give better estimate of the ground state energy of a localized 
proton.

The energy difference $\Delta E$, is a functional of two functions 
$\Psi \left( r\right) $ and $n\left( r\right) $. The physical constraint 
is that the variation of the neutron background conserves the baryon number. 
We should thus look for such functions $\Psi(r) $ and $n\left( r\right) $ 
that minimize the functional

\begin{equation}
f\left[ n\left( r\right) ,\Psi _P\left( r\right) \right] 
= \Delta E - \lambda 
\int_{V} \left[ n\left( r\right) - n_N \right]  d^3r - E \left[ 
\int_{V} \Psi_P^{*} \left( r\right) \Psi _P\left( r\right) 
d^3r - 1 \right] ,  \label{Functional}
\end{equation}

\ni where we explicitly include constraints of the baryon number conservation, 
Eq.(\ref{BaryonConserv}),
and the proton wave function normalization

\begin{equation}
\int_{V} \Psi _P^{*}\left( r\right) \Psi _P\left( r\right)  d^3r -1 =0.
\label{BaryonConserv}
\end{equation}

The Euler-Lagrange equations corresponding to the functional 
(\ref{Functional}) 
can be found easily. The differentiation with respect to $\Psi _{P}^{\ast }$ 
gives the Schr\"odinger equation for the proton impurity:

\begin{equation}
-{ 1 \over 2m_P} \nabla ^2\Psi _P\left( r\right) +\left[ 
\mu _P\left( n\left( r\right) \right) -\mu _P\left( n_N\right) \right] 
\Psi _P\left( r\right) = E_P\Psi _P\left( r\right) .
\end{equation}

\ni Differentiation with respect to $n\left( r\right) $ gives the second-order 
equation for the neutron density distribution $n(r)$:

\begin{equation}
{\partial \mu _P\left( n\left( r\right) \right) \over \partial n\left( 
r\right) }
\Psi _P^{*}\left( r\right) \Psi _P\left( r\right) +\mu _N\left(
n\left( r\right) \right)  + 2B_N {d^2n\left( r\right) \over dr^2} 
- \lambda =0 .  \label{Diff by n}
\end{equation}

\ni The boundary conditions the functions $\Psi \left( r\right)$ and $n\left( 
r\right) $ 
obey at $ r \rightarrow \infty $ are: $n\left( r\right)  = n_N$, 
$\left| \Psi _P\left( r\right) \right|^2=0 $. This allows us to identify 
the Lagrange multiplier $\lambda $ with the neutron chemical potential,

\begin{equation}
\lambda =\mu _N\left( n_N\right).
\end{equation}

To calculate the cell energy we adopt for the proton wave function 
the Gaussian form used in the previous section and solve with this ansatz 
the equation (\ref{Diff by n}). The rms radius of the proton probability 
distribution, 
$R_P$, is treated as a variational parameter. Numerical solutions of 
eq.(\ref{Diff by n})
are presented in Figs.14 and 15, where we show the neutron density 
distributions $n(r)$
obtained from equation (\ref{Diff by n}). As one can notice the neutron 
background has 
somewhat different shape than that used in Sect.3. In particular, 
at higher mean neutron densities $n_N$ there appears a significant density 
enhancement 
at the well boundary which considerably strengthens the localization effect. 
With the simple method of Sect.3 the neutron distribution around the proton is 
a monotonically increasing function of the radius.

Results of calculations of the energy difference $\Delta E$ with the 
self-consistent method are presented in Figs.3 and 4 as curves labeled with 
the value of the neutron matter density only. In Figs.5 - 10  the proton 
contribution and all components of the neutron background contribution to 
the full energy difference are shown together with those obtained in Sect.3.
The neutron  background energy, Figs. 7 and 8,  calculated with the 
self-consistent method is
well below the simple estimate of Sect.3 for low values of the proton 
distribution radius $R_P$. The reduction of the energy is even bigger for the 
gradient term contribution, Figs.9 and 10. An opposite effect is observed for 
the proton energy contribution, Figs. 5 and 6, where the energy corresponding 
to the self-consistent method is higher than that calculated with simple trial
functions in Sect.3.

\section{Conclusions and implications}

Our self-consistent method gives lower energies of localized protons 
than the variational method used in sect.3. In Figs.3 and 4 we compare 
the energy difference $\Delta E$ obtained with both methods, 
for the MS and FPR models of nuclear interactions. As functions of 
the radius $R_P$ 
the energy differences $\Delta E$ for the old variational method and for 
the selfconsistent method depart from one another only for small values of 
$R_P$. 
At $R_P>1 fm$ the curves for both methods in Figs.3 and 4 are practically 
identical. 
However at small $R_P$ the self-consistent method gives significantly lower 
energies. 
The minimum values of $\Delta E$ are considerably below those found with 
the old method and they occur at somewhat smaller radii.
Our study indicates also that with the growing curvature coefficient 
the localization density grows. The results presented in Fig.11 show that this 
growth  is  weak and the values corresponding to both methods are quite 
similar. Also, the dependence of the localization density on the proton 
effective mass, Fig.13 is very similar for both methods.

To conclude, the self-consistent calculations improve the estimate 
of the energy of the cell containing a localized proton, especially at small 
values of the rms proton  radius $R_P$. The proton contribution  $E_P$  and 
the gradient term contribution  to $\Delta E$ are most affected by the new 
method. The ultimate goal 
is to calculate the proper wave function of the proton, which would give 
the true energy of the localized state.

Results of our calculations for nuclear interactions we use indicate that 
the proton impurity in neutron star matter becomes localized at densities 
above 
$0.5-1.0 fm^{-3}$. The selfconsistent method gives  lower energies of 
localized 
protons 
and smaller threshold localization densities than simple variational 
method 
with trial functions. This has important consequences for neutron stars as 
densities in this range correspond to inner core of neutron stars with masses 
exceeding one solar mass, $M> 1M_{\odot}$. In Fig. 16 we show neutron star 
masses corresponding to all nuclear interactions used in the calculations 
reported above.

\newpage 

\rysunek{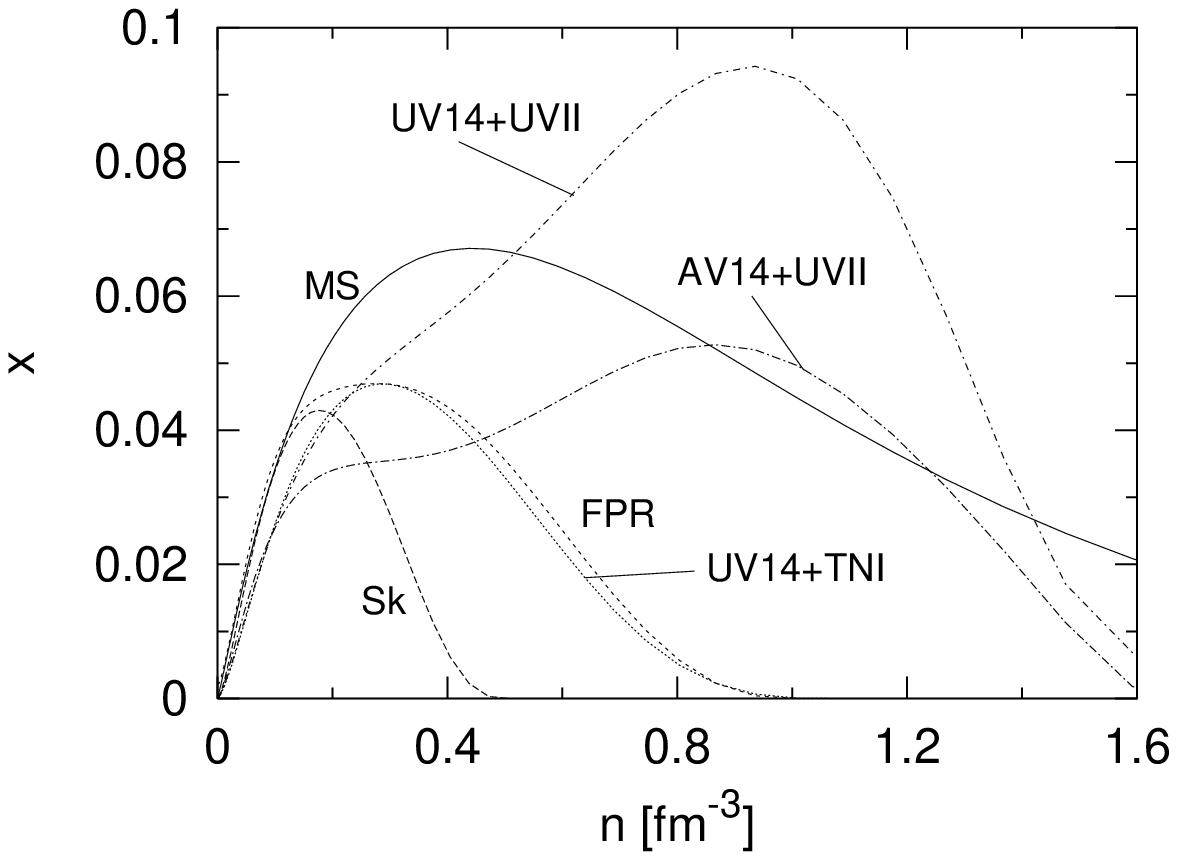}{The proton fraction of the neutron star matter as a function 
of baryon number density for indicated nuclear interaction models.}
\rysunek{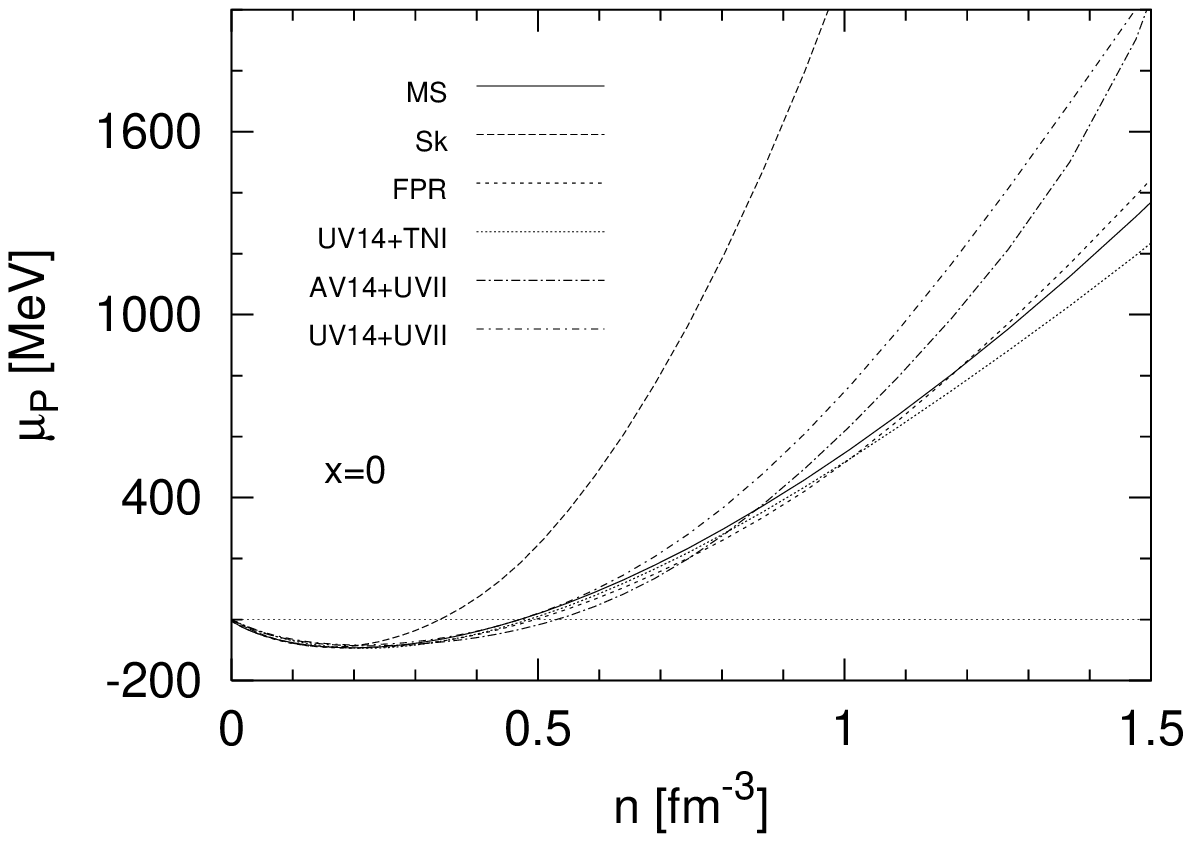}{The proton chemical potential in pure neutron matter 
as a function of
baryon number density for the same interactions as in Fig.1.}
\rysunek{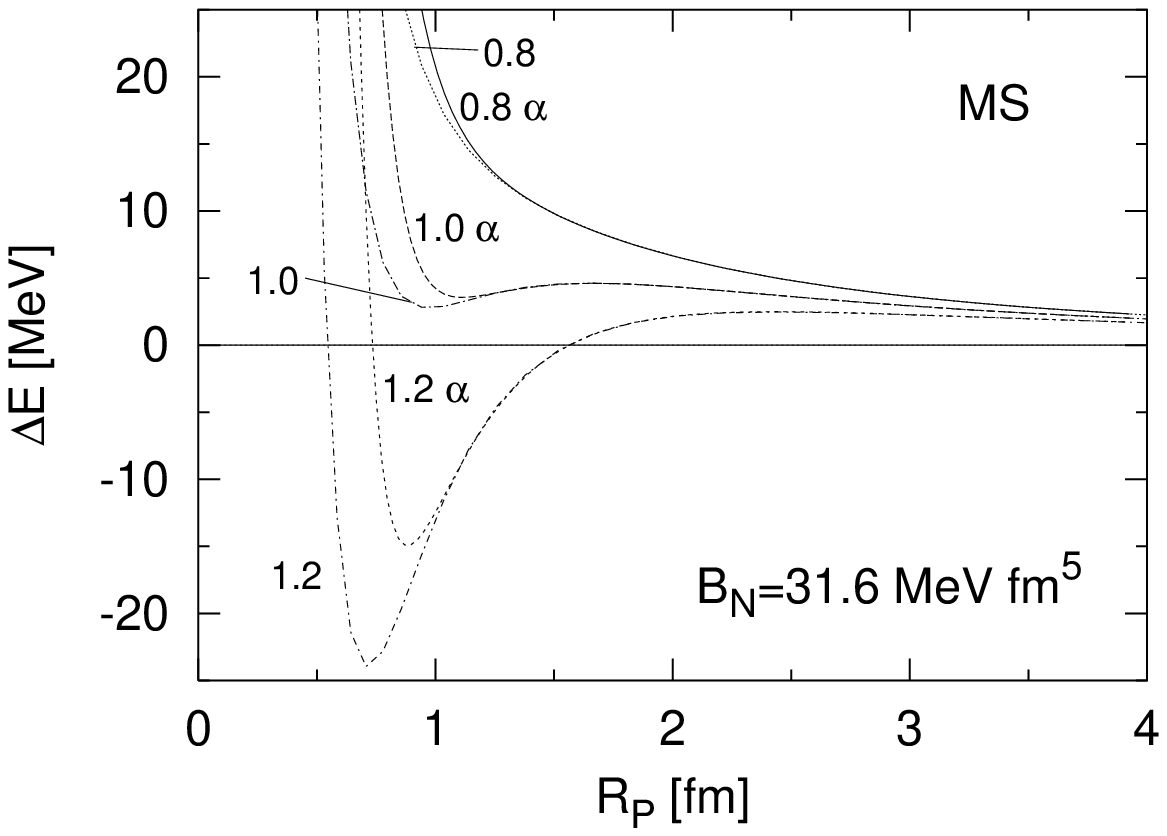}{The energy difference $\Delta E$ as a function of the 
proton 
rms radius for the Myers and Swiatecki interaction. The curves corresponding 
to the self-consistent calculations are labeled with the value of the neutron 
matter density in $[fm^{-3}]$. The curves labeled additionally with the letter 
$\alpha$ correspond to the simple method of Sect.3.}
\rysunek{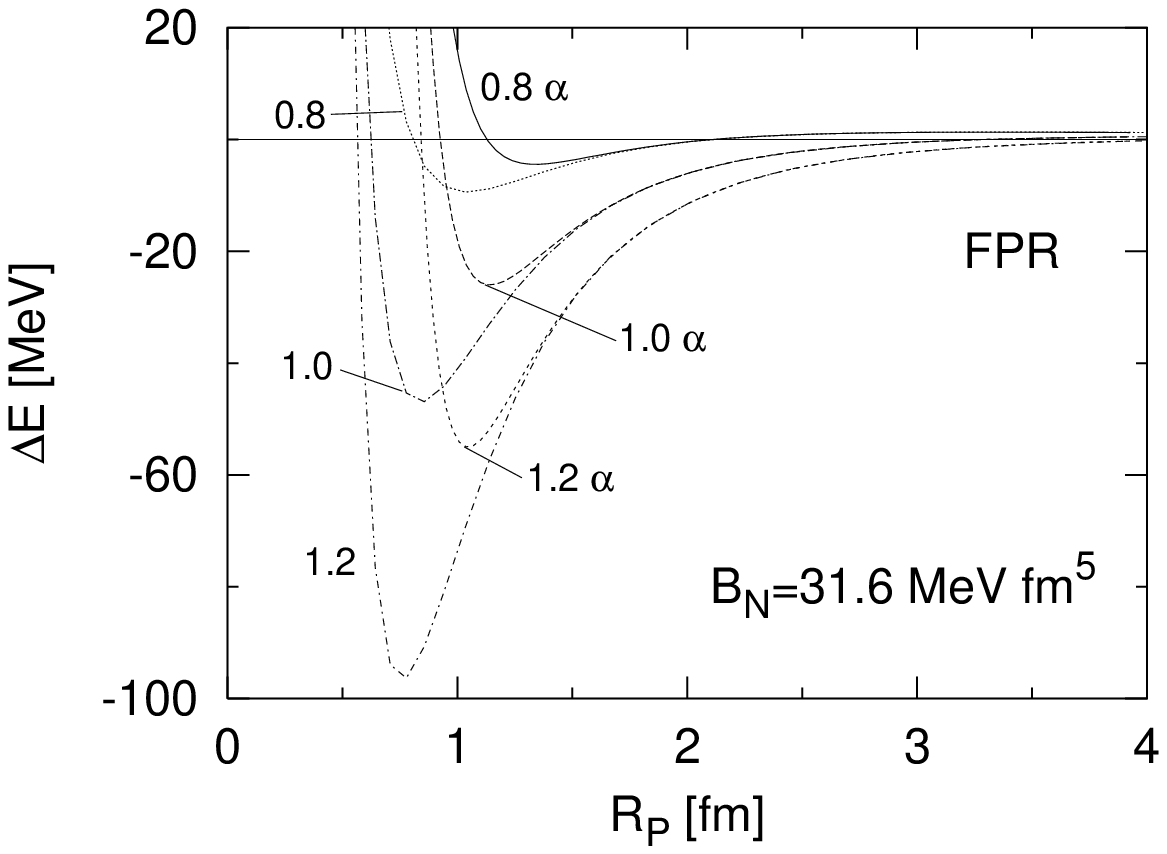}{The same as in Fig.3  
for the Friedman-Pandharipande-Ravenhall nuclear interaction.}
\rysunek{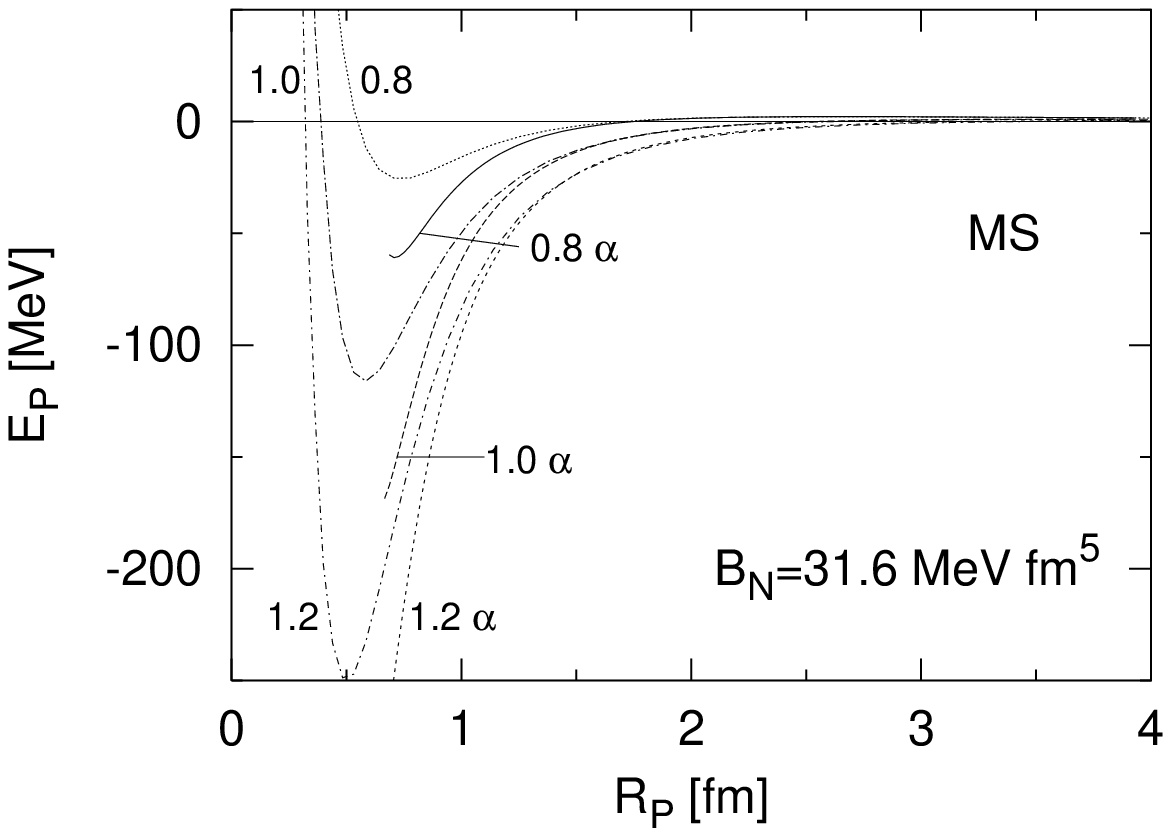}{The proton contribution to $\Delta E$ for 
the Myers and Swiatecki model of nuclear interactions. Curves labeled as in 
Fig.3.}
\rysunek{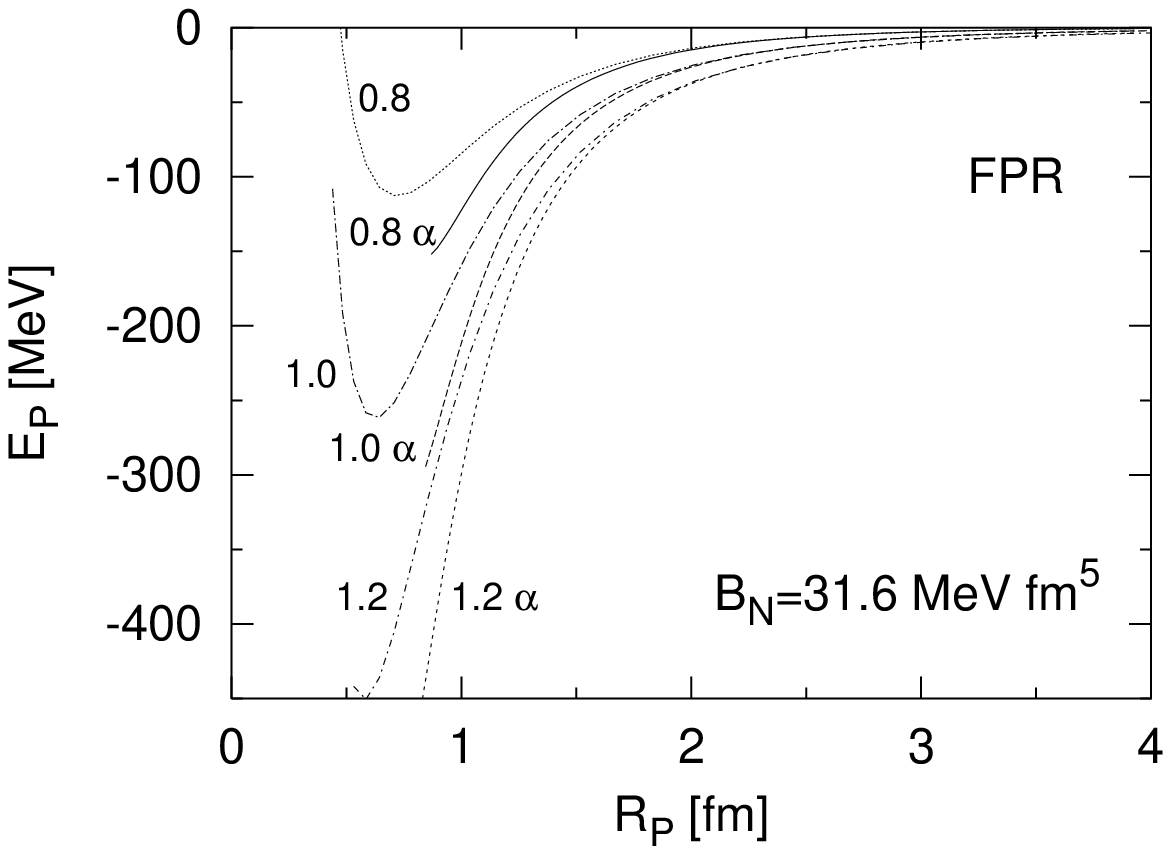}{The same as in Fig.5 for 
the Friedman-Pandharipande-Ravenhall interactions.}
\rysunek{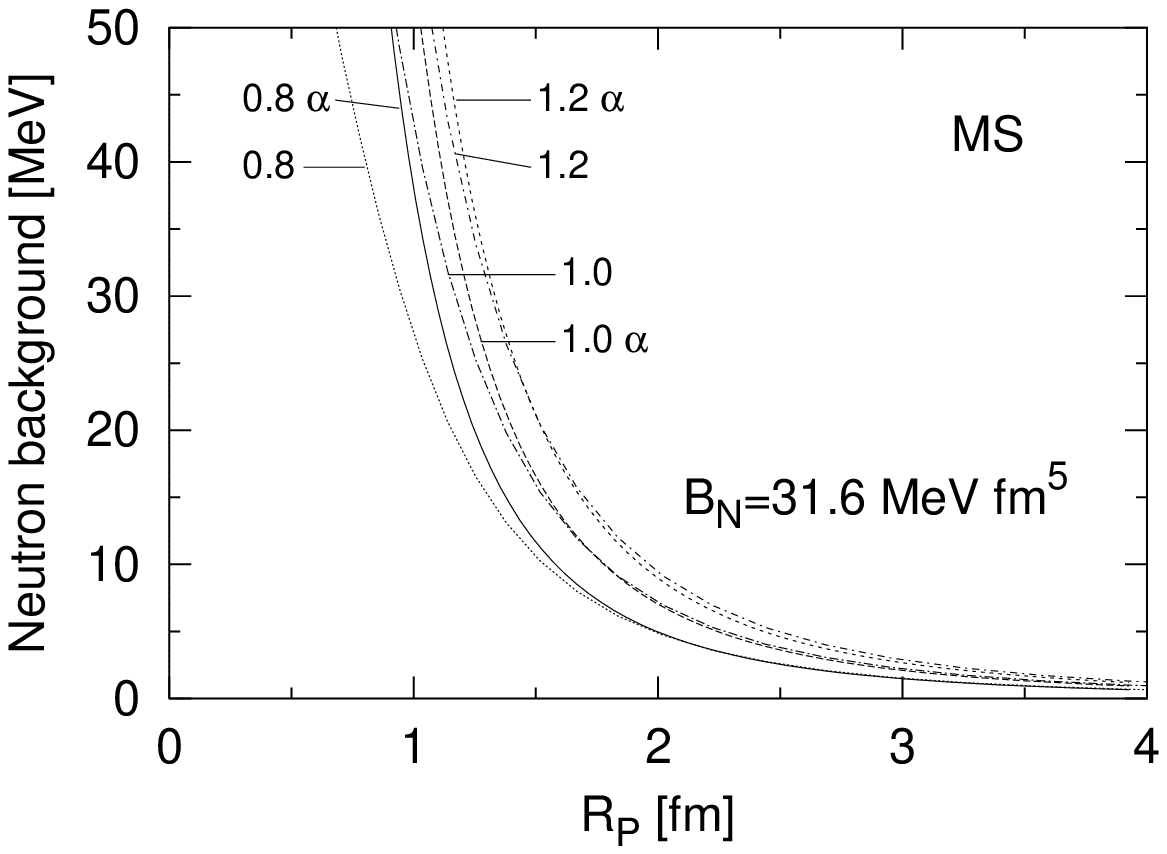}{The neutron background contribution to $\Delta E$
for the Myers and Swiatecki model of nuclear interactions. Curves labeled as 
in 
Fig.3}
\rysunek{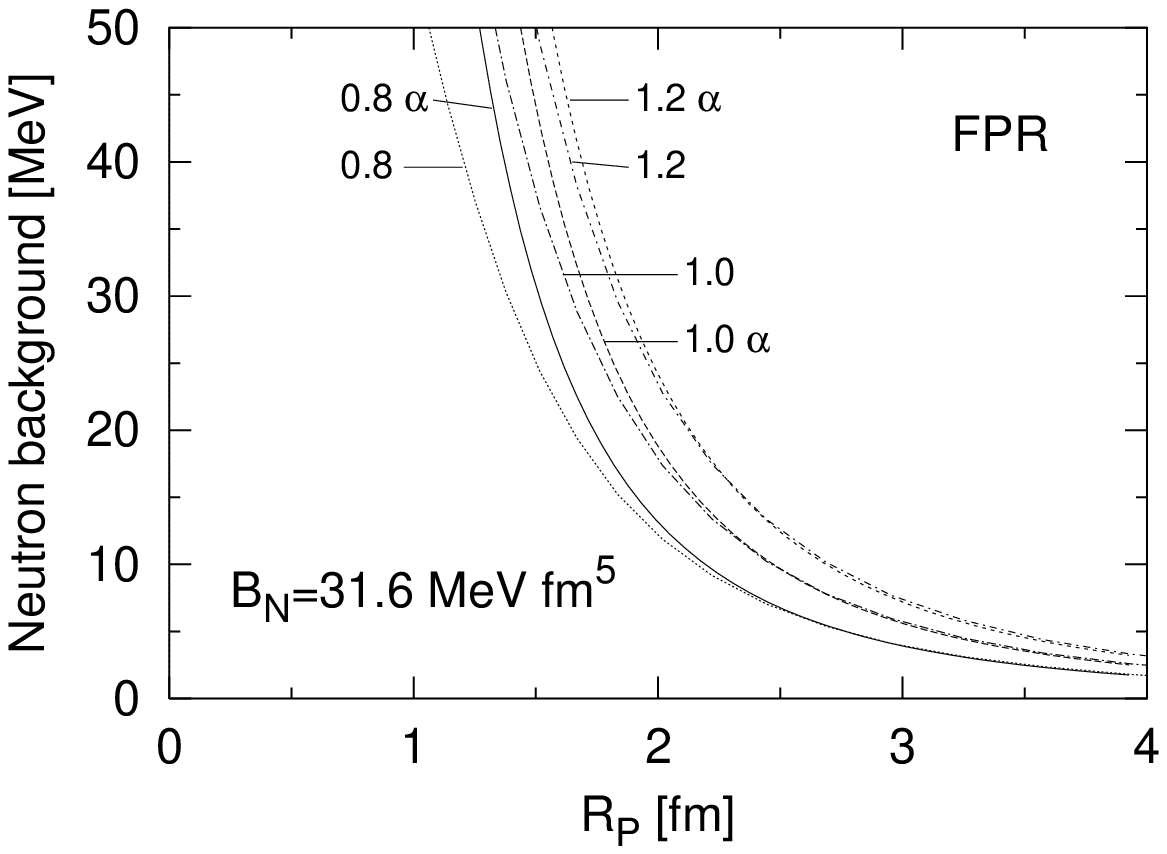}{The same as in Fig.7
for the Friedman-Pandharipande-Ravenhall interaction.}
\rysunek{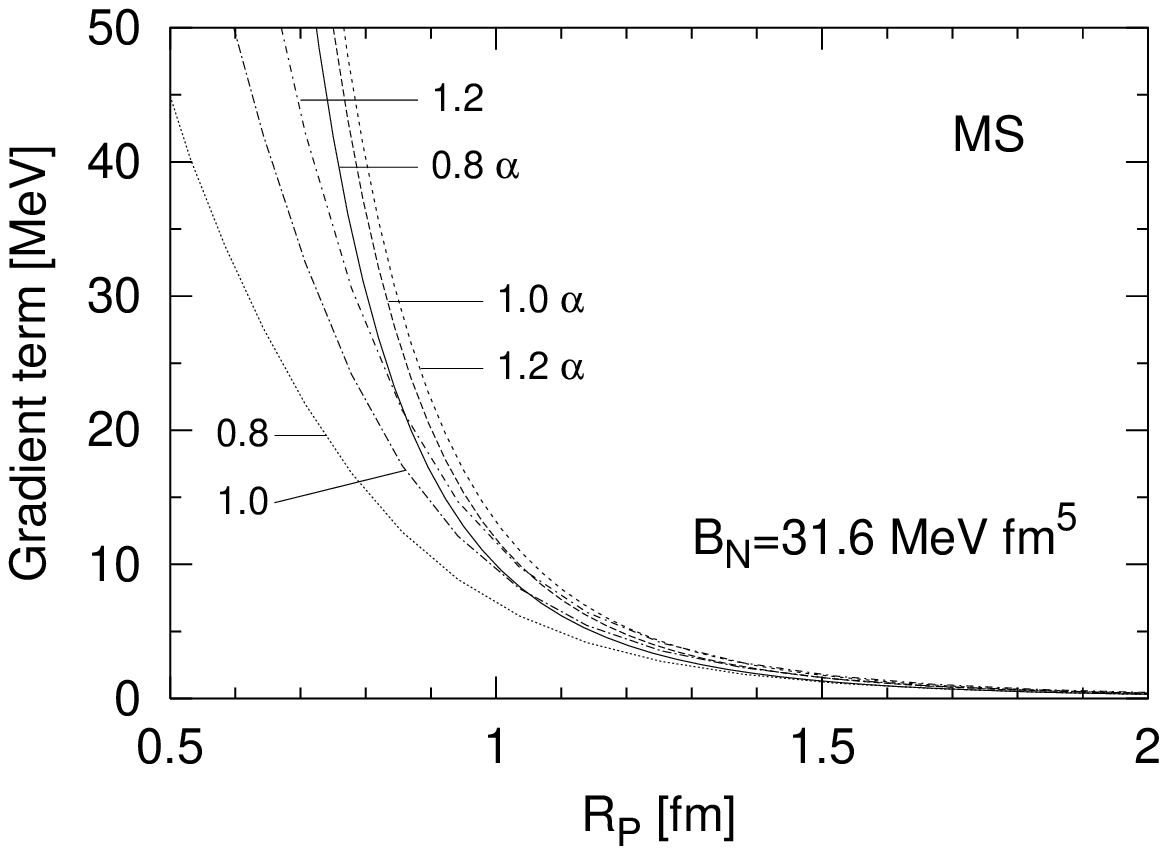}{The gradient term contribution to $\Delta E$ for the 
Myers and Swiatecki nuclear interactions. Curves labeled as in Fig.3}
\rysunek{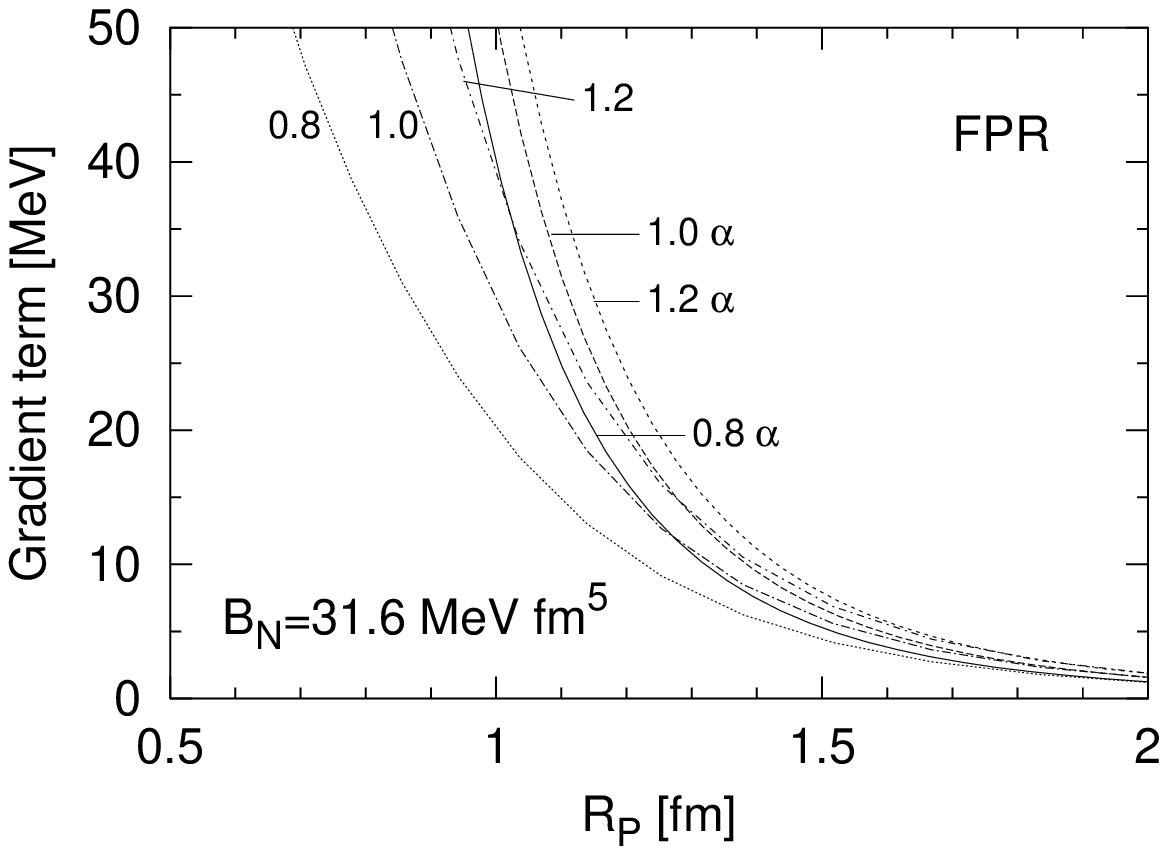}{The same as in Fig.9 
for the Friedman-Pandharipande-Ravenhall interactions.}
\rysunek{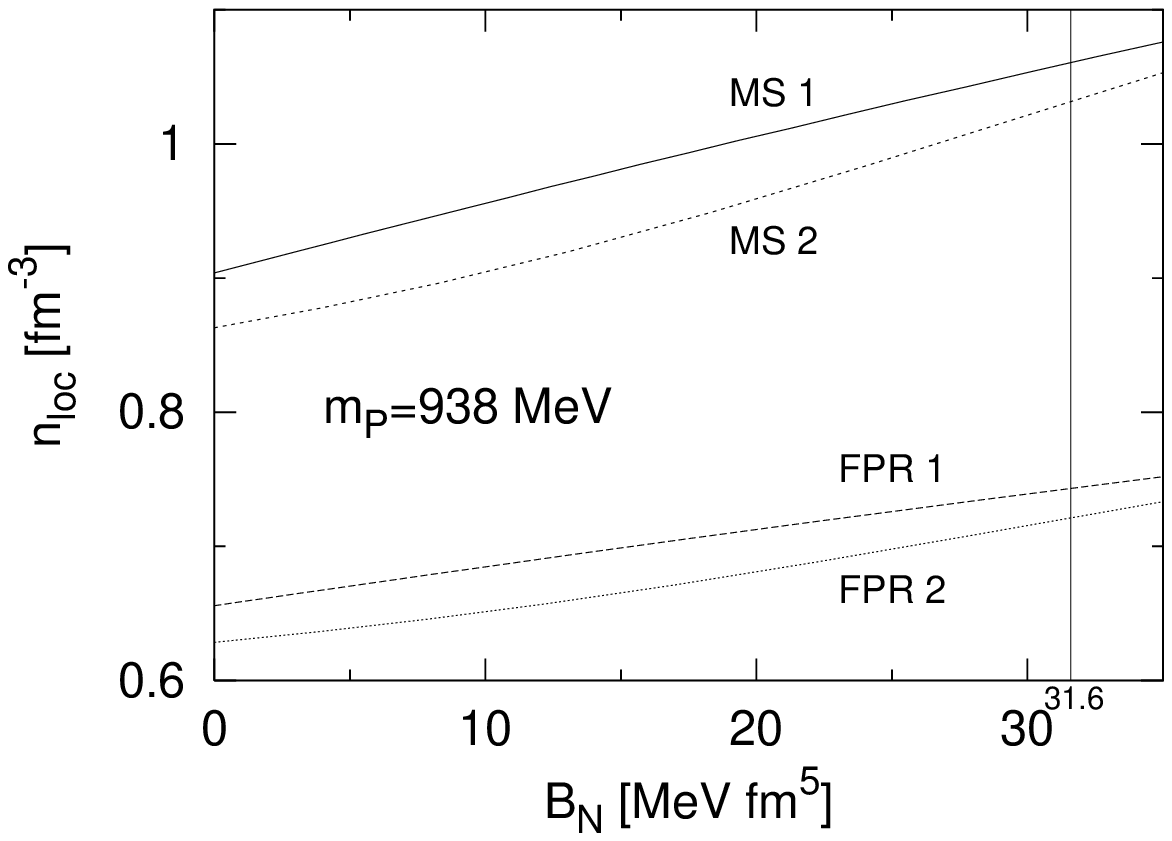}{The threshold density for proton localization versus
the
curvature coefficient $B_N$. The curves MS1 and FPR1 correspond to the simple 
method of Sect.3. The curves MS2 and FPR2 correspond to the self-consistent 
calculations.}
\rysunek{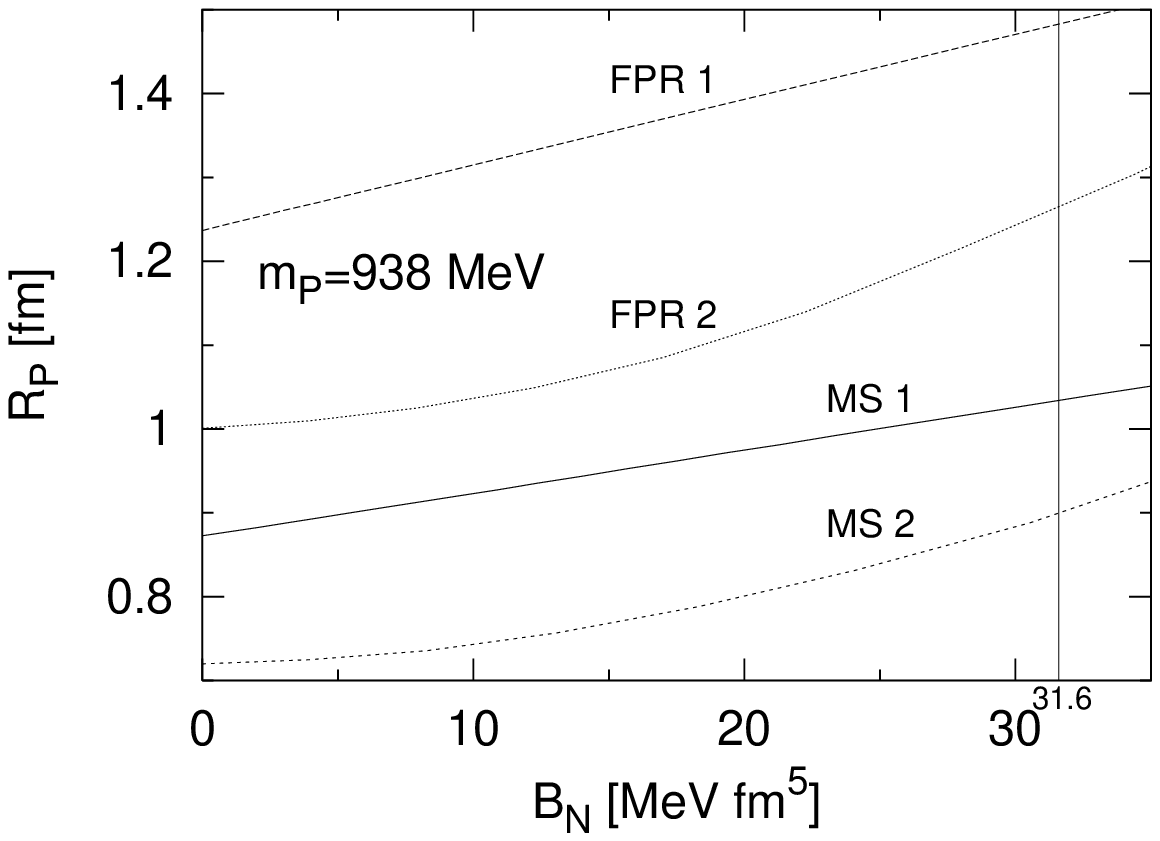}{The rms proton distribution radius at the threshold 
density as a function of
the curvature coefficient $B_N$. Curves labeled as in Fig.11}
\rysunek{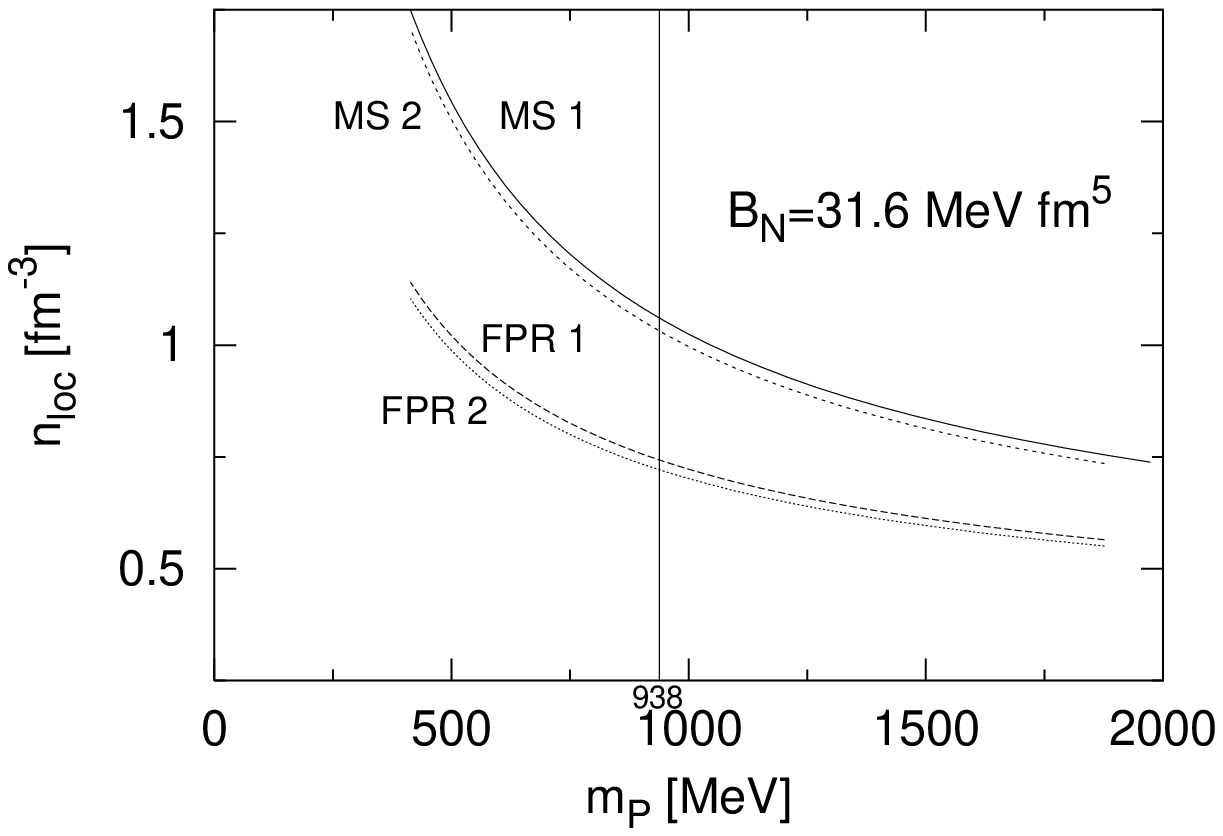}{The threshold density for proton localization as a 
function of
the proton effective mass. Curves labeled as in Fig.11}
\rysunek{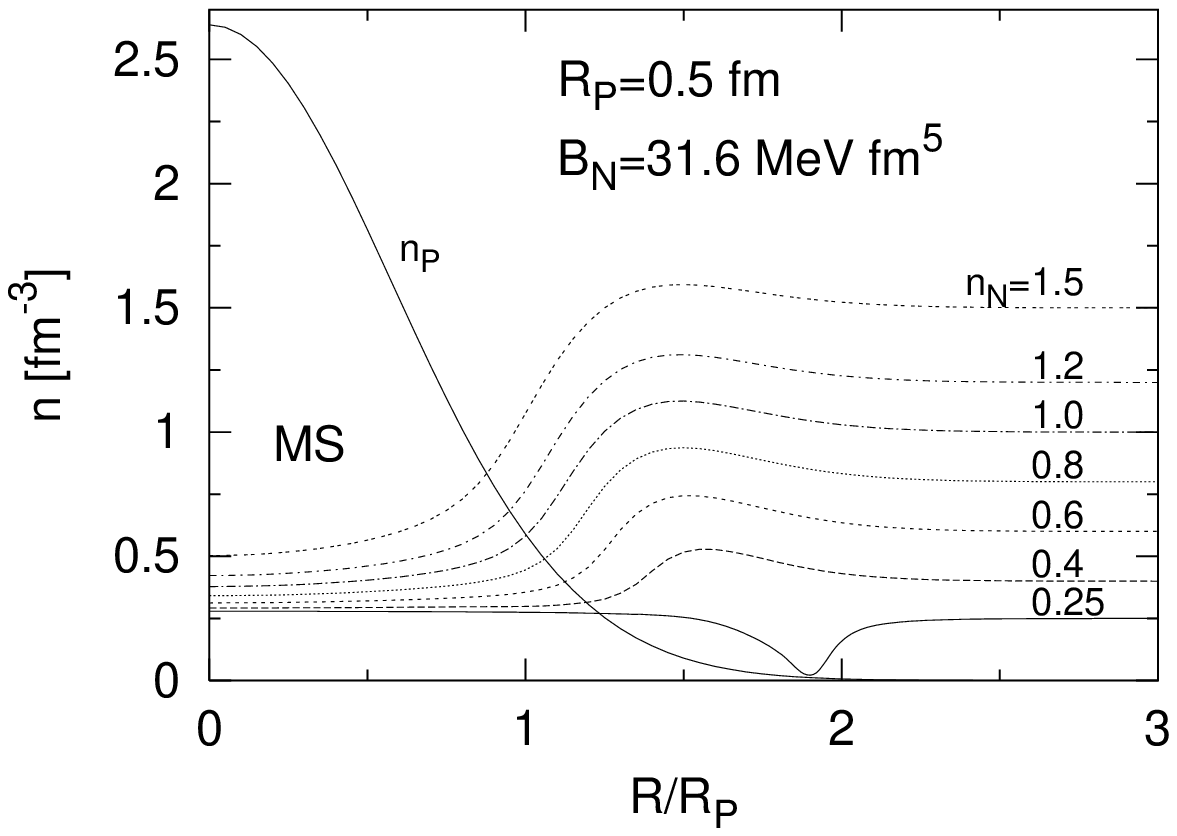}{The neutron density distribution obtained from Eq.(16) 
for indicated neutron 
matter 
densities (in $[fm^{-3}]$) for the Myers and Swiatecki interactions. The 
localized proton distribution $n_P$ is also shown.}
\rysunek{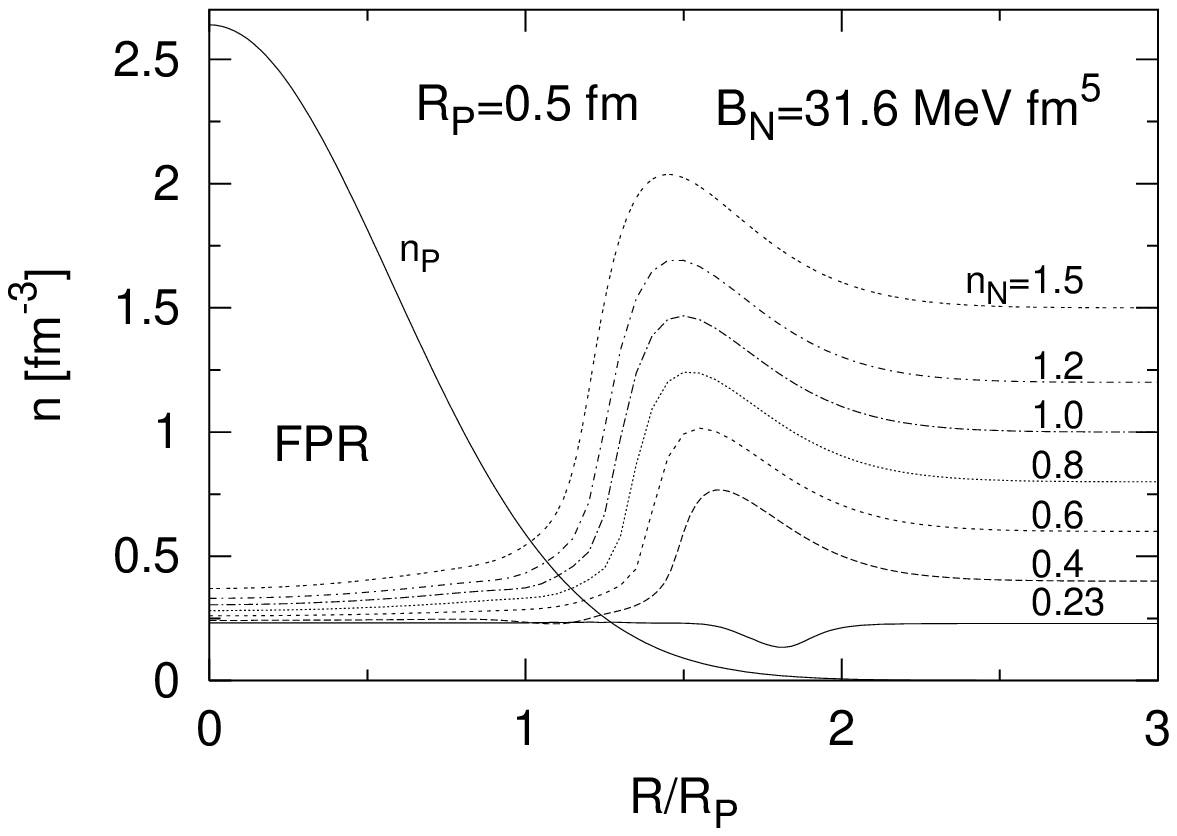}{The same as in Fig.14   for the 
Friedman-Pandharipande-Ravenhall  
interactions.}
\rysunek{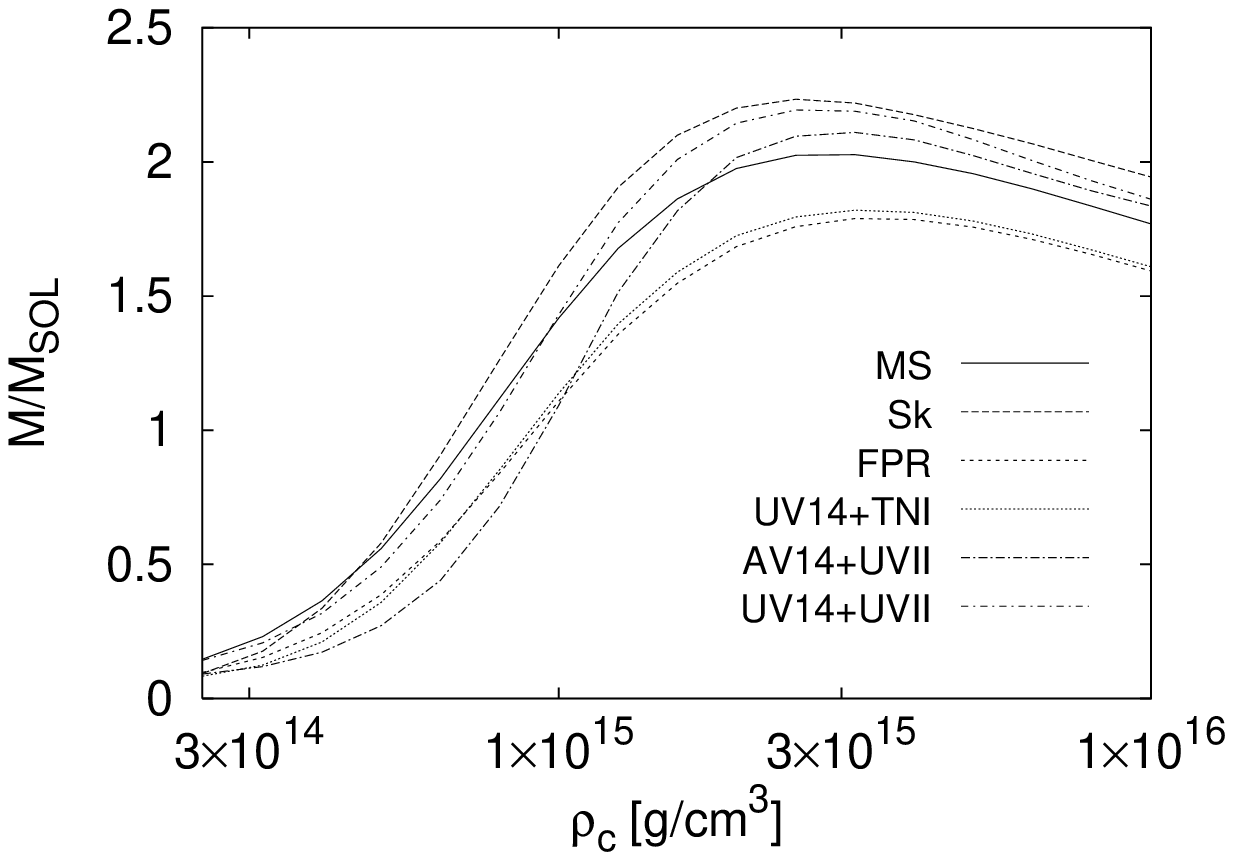}{Neutron star masses (in solar units) as functions of 
the central density for nuclear interaction models from Fig.1.}

\end{document}